\documentclass[%
 reprint,
superscriptaddress,
 amsmath,amssymb,
 aps,
floatfix,
onecolumn,
prl,
notitlepage,
longbibliography
]{revtex4-1}
\pdfoutput=1
\usepackage{graphicx}
\usepackage{dcolumn}
\usepackage{bm}

\begin{document}

\preprint{APS/123-QED}

\title{Particle shapes leading to Newtonian dilute suspensions}


\author{David A. King} \email{dak43@cam.ac.uk}
\affiliation{%
Cavendish Laboratory, University of Cambridge, J. J. Thomson Ave., Cambridge CB3 0HE, UK
}%
\author{Masao Doi}%
\affiliation{%
Centre of Soft Matter and its Applications, Beihang University, Beijing 100191, China 
}%
\author{Erika Eiser} \email{ee247@cam.ac.uk}
\affiliation{%
Cavendish Laboratory, University of Cambridge, J. J. Thomson Ave., Cambridge CB3 0HE, UK
}%

\date{\today}

\begin{abstract}
It is well known that suspensions of particles in a viscous fluid can affect the rheology significantly, producing a pronounced non-Newtonian response even in dilute suspension. However, it is unclear \textit{a priori} which particle shapes lead to this behaviour. We present two simple symmetry conditions on the shape which are sufficient for a dilute suspension to be Newtonian for all strain sizes and one sufficient for Newtonian behavior for small strains. We also construct a class of shapes out of thin, rigid rods not found by the symmetry argument which share this property for small strains. 
\end{abstract}

\maketitle

The theoretical study of rigid particle suspensions dates back to Einstein's seminal work \cite{Einstein1906EineMolekul-dimensionen, Einstein1911BerichtigungMolekul-dimensionen}. He showed that introducing a dilute suspension of spheres to a viscous solvent increased the solvent's viscosity from $\eta_s$ to $\eta_s(1 + 2.5 n \nu)$, where $n$ is the number density and $\nu$ is the volume of each sphere. Since then, a wide variety of systems have been successfully modelled, from polymeric fluids \cite{Doi1986TheDynamics} to active bacterial suspensions \cite{Saintillan2018RheologyFluids}.
\begin{figure}
\includegraphics[width=12cm]{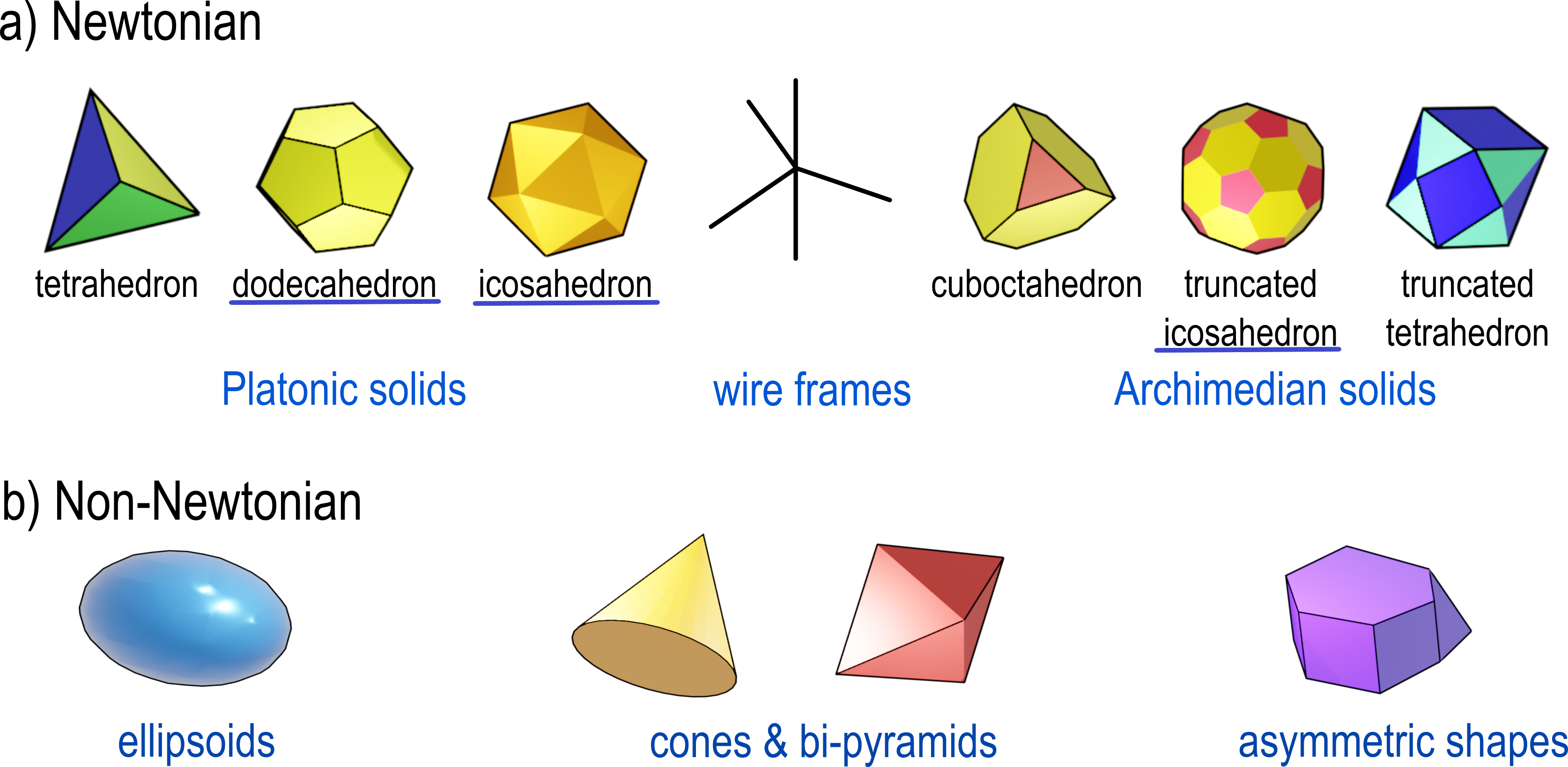}
\caption{\label{fig:shapes}a) Examples of particle shapes which have Newtonian dilute suspensions for small strains. Due to their symmetry, the Platonic and Archimedean solids produce no elasticity. Some wire frame shapes share this property if the ratios of the rod lengths are chosen correctly. The underlined shapes lead to Newtonian suspensions for all strain sizes. b) Examples of particle shapes which, generally, lead to non-Newtonian dilute suspensions.}
\end{figure}

The rheology of a fluid is understood through the constitutive equation relating the stress, $\sigma$, and the rate of strain tensor $\kappa$. For a Newtonian fluid, the stress is linearly related to the strain by a time independent $4^{th}$ rank tensor. For isotropic, incompressible fluids this tensor is isotropic and determined solely by the fluid's viscosity. For small strains, the stress in a general fluid is characterised by the complex viscosity, $\eta^{*}(\omega)$. This is defined such that the stress response to the oscillatory shear, $\kappa_{xy} = \dot{\gamma} \Re (e^{i \omega t})$, is given by $\sigma_{xy} =  \dot{\gamma} \Re (\eta^{*}(\omega) e^{i \omega t})$ where $\dot{\gamma}$ is the shear rate and $\omega$ is the angular frequency. The real and imaginary parts of $\eta^{*}$ represent the viscosity and elasticity of the solution respectively, both of which may be dynamic. A fluid whose complex viscosity has both real and imaginary parts is viscoelastic. The complex viscosity of a Newtonian fluid is a real constant, independent of frequency. 

The presence the particles can drastically change the behaviour of a fluid, producing a pronounced non-Newtonian response. While the most extreme phenomena occur at large concentrations or strains, it has been shown that dilute suspensions of certain particle shapes also lead to viscoelastic behaviour for small strains. This is not the case for spheres. Notably, when their Brownian motion is taken into account, rigid rod-like particles show a finite, linear elastic response in dilute suspension \cite{Doi1986TheDynamics}. Other particle shapes, such as spheroids \cite{Hinch1972TheParticles} and propeller-like particles \cite{Makino2004ViscoelasticityShape}, have also been shown to behave similarly. Clearly, the presence of elasticity in these suspensions depends on the particles' shape. While the methods for finding the stress in such systems are well known \cite{Kim2005MicrohydrodynamicsApplications,Happel1973LowMedia.}, it is unclear \textit{a priori} which particle shapes will lead to a non-Newtonian response, without simulations or cumbersome calculations.

In this letter, we aim to understand and characterise the properties that a particle shape must have for a dilute suspension to be Newtonian. We determine two simple symmetry conditions on the particle shape which, when both are satisfied, are sufficient for the suspension to be Newtonian. These are derived by considering the symmetries of the particle, without referencing its specific shape. As long as the particles' positions remain uniformly distributed and inter-particle interactions are negligible, these hold for all strain sizes. In the case of small strains, one of these two conditions is relaxed, and there is only one sufficient condition for purely viscous behaviour. Examples of particle shapes with Newtonian dilute suspensions are shown in Fig.(\ref{fig:shapes}a), the underlined shapes have a Newtonian response for all strain sizes. The conditions presented here makes this identification straightforward where explicit calculations would be extremely difficult, e.g. for the Archimedian solids.

The ability to predict the presence of a non-Newtonian response for particles with arbitrary shapes has become highly relevant given the emergence of sophisticated techniques \cite{Seeman1982NucleicLattices,Rothemund2006FoldingPatterns} to design the  shape of nano-particles. We specifically reference DNA nanostars \cite{Bomboi2019Cold-swappableGels, Biffi2013PhaseNanostars, Xing2018MicrorheologyHydrogels}, constructed from linked double stranded DNA sequences, which have recently been synthesised and studied for their potential biomedical and nano-engineering applications. 

We provide a general method for describing the rheology of dilute suspensions of these particles, based on the Onsager Principle \cite{Onsager1931ReciprocalI., Onsager1931ReciprocalII.}, a powerful tool for describing the behaviour of a wide range of systems \cite{Doi2011OnsagersMatter, Doi2013SoftPhysics}. We use this approach to demonstrate the predictions of the symmetry argument for small strains for specific shapes, and construct a class of shapes not found by the symmetry argument whose dilute suspensions also have a purely viscous linear response.

\paragraph{General Shapes - Background:}\label{sec:generalshapes}
We begin with a brief overview of the method for determining the viscoelasticity of a dilute suspension of particles of general shape, a full description may be found in Makino and Doi \cite{Makino2004ViscoelasticityShape}.

There are two contributions to the stress in these systems: one arising purely from the hydrodynamics of the suspended particles and the other from their Brownian motion. We assume that the particle sizes, velocity and viscosity of the fluid are such that the Reynolds number may be taken to be small. To describe the hydrodynamics in this regime, we only need to consider one particle whose linear velocity, $\textbf{v}$, and angular velocity, $\bm{\Omega}$, are linearly related to those of the fluid and the rate of strain tensor of the flow via mobility matrices which depend on the geometry and orientation of the particle \cite{Kim2005MicrohydrodynamicsApplications, Happel1973LowMedia.}. 

The Brownian motion of the particle is included through the effective potential $U_{B} = k_{B} T \log \psi$, written in terms of the distribution function of the particle, $\psi$. The orientation of the particle is represented using a right-handed, ortho-normal set of three vectors  $\textbf{u}_1$,$\textbf{u}_2$ and $\textbf{u}_3$ fixed to the particle. The viscoelastic properties are calculated by assuming no external forces and that $\psi$ is independent of the position of the particle. 

Using this framework, the hydrodynamic stress per-particle is,
\begin{equation}
\label{stresshyd}
(S_H)_{ij} = K_{ijkl} \kappa_{kl},
\end{equation}
and the Brownian stress per particle is,
\begin{equation}
\label{stressb}
(S_{B})_{ij} = h_{ijk} \mathcal{R}_k (-k_{B} T \log \psi),
\end{equation}
where $K$ and $h$ are mobility tensors and in the above equations and henceforth all indices used indicate the laboratory frame components, and repeated indices are summed unless otherwise stated. We also used the rotational derivative operator, $\bm{\mathcal{R}} = \textbf{u}_{\mu} \times \partial / \partial \textbf{u}_{\mu}$, where the index $\mu = 1,2,3$ is summed. 

In general the stress tensor for the system is written as:
\begin{equation}
\label{stressgen}
\sigma = - p \mathbb{I} + 2 \eta_s \kappa - n \langle S_B + S_H\rangle 
\end{equation}
where $p$ is the pressure, $\mathbb{I}$ the $3 \times 3$ identity matrix and $\eta_s$ the solvent viscosity. The angle brackets denote averaging over the orientational distribution $\langle \cdots \rangle = \int d\textbf{u}_1 d\textbf{u}_2 \ \psi \ (\cdots) $. All viscoelastic properties follow from these expressions.

\paragraph{Symmetry Conditions:} \label{symcondition}
 A suspension of particles can be Newtonian if the Brownian and hydrodynamic stresses satisfy certain conditions. In the linear regime, the hydrodynamic stress contributes a constant to the complex viscosity and is therefore always a purely viscous, Newtonian contribution. The Brownian stress depends implicitly on the strain, via $\psi$, and can be shown to decay with time, contributing both real and imaginary parts to the complex viscosity. Therefore, in the linear regime, if the Brownian stress vanishes, the solution is purely viscous and Newtonian. Since the Brownian stress tensor is traceless, if it is isotropic then it must vanish identically and the suspension is purely viscous.

For larger strains the Hydrodynamic stress can produce non-Newtonian stresses. However, if the mobility tensor, $K$, is independent of the particle's orientation then its average must be independent of time and therefore $\langle S_H \rangle$ is purely viscous. For $K$ to be independent of the particles orientation in the laboratory frame it must be an isotropic tensor, so that it is invariant under all rotations from one orientation to another. 

We therefore have two conditions which, if both are satisfied, are sufficient for the rheology to be Newtonian for all strain sizes: I) the Brownian Stress tensor must be isotropic and II) the mobility tensor $K$ must be isotropic. These statements are true assuming the particle distribution remains uniform and inter-particle interactions are negligible. If we only require the suspension to be purely viscous in the linear regime, then only the first condition needs to hold. 

To determine which particle shapes satisfy these conditions we consider the symmetry group, $\mathcal{G}$, of the particle. This is a set of $3 \times 3$ rotation matrices, $R$, which leave the orientation of the particle unchanged. For example, consider a cubic particle whose faces and edges are all identical and whose density is uniform. If we rotate this particle by $\pi/2$ about an axis passing through the centre of any one of its faces, the orientation of the particle is outwardly the same. The only difference is the definition of the unit vectors, $\textbf{u}_{1,2,3}$. 

Under the action of $R$, the vectors' components transform according to the standard rule,
\begin{equation}
\label{unitvectrans}
    u^{\mu}_{i} \to (u')^{\mu}_{i} = R_{ai} u^{\mu}_{a},
\end{equation}
where the symbol $u^{\mu}_{i}$ denotes the $i^{th}$ component of the vector $\textbf{u}_{\mu}$ \footnote{Note that the vector remains of unit magnitude, since the determinant of the rotation is one}. 

The physical consequence of these symmetries is that by applying the above transformation to the orientation vectors and imposing the same background fluid flow, the response of the particle would be the same as measured in the laboratory frame. Specifically, $S_B$ and $S_H$ are preserved under the action of the transformation, \ref{unitvectrans}.

Beginning with $S_B$, we use equation (\ref{stressb}) and write it in terms of the transformed vectors $\textbf{u}'$,
\begin{equation}
(S_{B})_{ij} = - R_{ai} R_{bj} R_{ck} h_{abc} R_{dk} \mathcal{R}_{d} U_B,      
\end{equation}
where we have taken into account that $\mathbf{\mathcal{R}} U_B$ and $h$ transform as a vector \footnote{generally the rotational derivative should transform as a pseudo-vector to be consistent in the Brownian torque, however, since we take $|R|=1$, there is no difference in the transformation laws.} and third rank tensor respectively and $|R|=1$.

We may then contract the index, $k$, to find that the Brownian stress must satisfy,
\begin{equation}
\label{stressbtrans2}
(S_{B})_{ij} =  (\textbf{R}^{T} \cdot \textbf{S}_B \cdot \textbf{R})_{ij}.
\end{equation}
This must hold for each member of the shape's symmetry group, therefore the following commutation relations are implied:
\begin{equation}
\label{commutator}
    [\textbf{S}_B , \textbf{R}] = 0 \ \ ; \ \ \forall \ \textbf{R} \in \mathcal{G}.
\end{equation}
We now make use of Schur's First Lemma \cite{Schur1905NeueGruppencharaktere}, which states that if a matrix commutes with all members of an irreducible representation of a group, then it is proportional to the identity matrix. An irreducible representation is one where each member cannot be cast in block diagonal form by the same similarity transformation \cite{Hall2015LieRepresentations}. 

Condition I can now be understood as a condition on the particle shape; its symmetry group  \textit{must} have an irreducible representation in $3 \times 3$ matrices. If it does, the relations, (\ref{commutator}), and Schur's First Lemma imply $\textbf{S}_B \propto \mathbb{I}$.

To ensure that a dilute suspension of particles is Newtonian for all strain sizes, condition II must be met. The relation between the Hydrodynamic stress and the strain via $K$ is mathematically the same as that between the stress, strain and elastic modulus for solids, with $K$ playing the role of the elastic modulus. The symmetry conditions needed for the solid elastic modulus to be isotropic have been widely studied \cite{Dresselhaus2008GroupMatter, Dresselhaus1991OnModuli}. By considering how the symmetries reduce the number of independent components of the tensor $K$, it has been shown that when $K$ is invariant under a symmetry group with an irreducible representation of degree five it is isotropic \cite{Dresselhaus1991OnModuli}. This is true for spheres, icosahedra and higher symmetry shapes, but not for cubes or lower symmetry shapes. Since shapes with icosahedral symmetry also satisfy condition I, they will be purely viscous and Newtonian for all strain sizes, whereas shapes with cubic or tetrahedral symmetry will only have this property in the small strain regime. 

We can now understand why a dilute suspension of spheres is purely viscous, whereas rigid, rod-like particles have a viscoelastic linear response in dilute suspension. The symmetry group of a sphere, $O(3)$, has a well known irreducible representation in $3 \times 3$ matrices, hence the elasticity \textit{must} vanish. Rods, on the other hand, are rotationally symmetric about their axis, chosen to align with $\textbf{u}_3$, and are symmetric under the inversion, $\textbf{u}_3 \to -\textbf{u}_3$. These symmetries do not have an irreducible representation in $3 \times 3$ matrices. Therefore, dilute suspension of rods \textit{can} have a finite viscoelasticity. 

Fig.(\ref{fig:shapes}a) shows examples of shapes that produce purely viscous dilute suspensions for all strain sizes and small strains. Next, we construct a particular class of shapes not found by this symmetry argument whose dilute suspensions have a purely viscous stress response to small strains.

\begin{figure}
{\includegraphics[width=12cm]{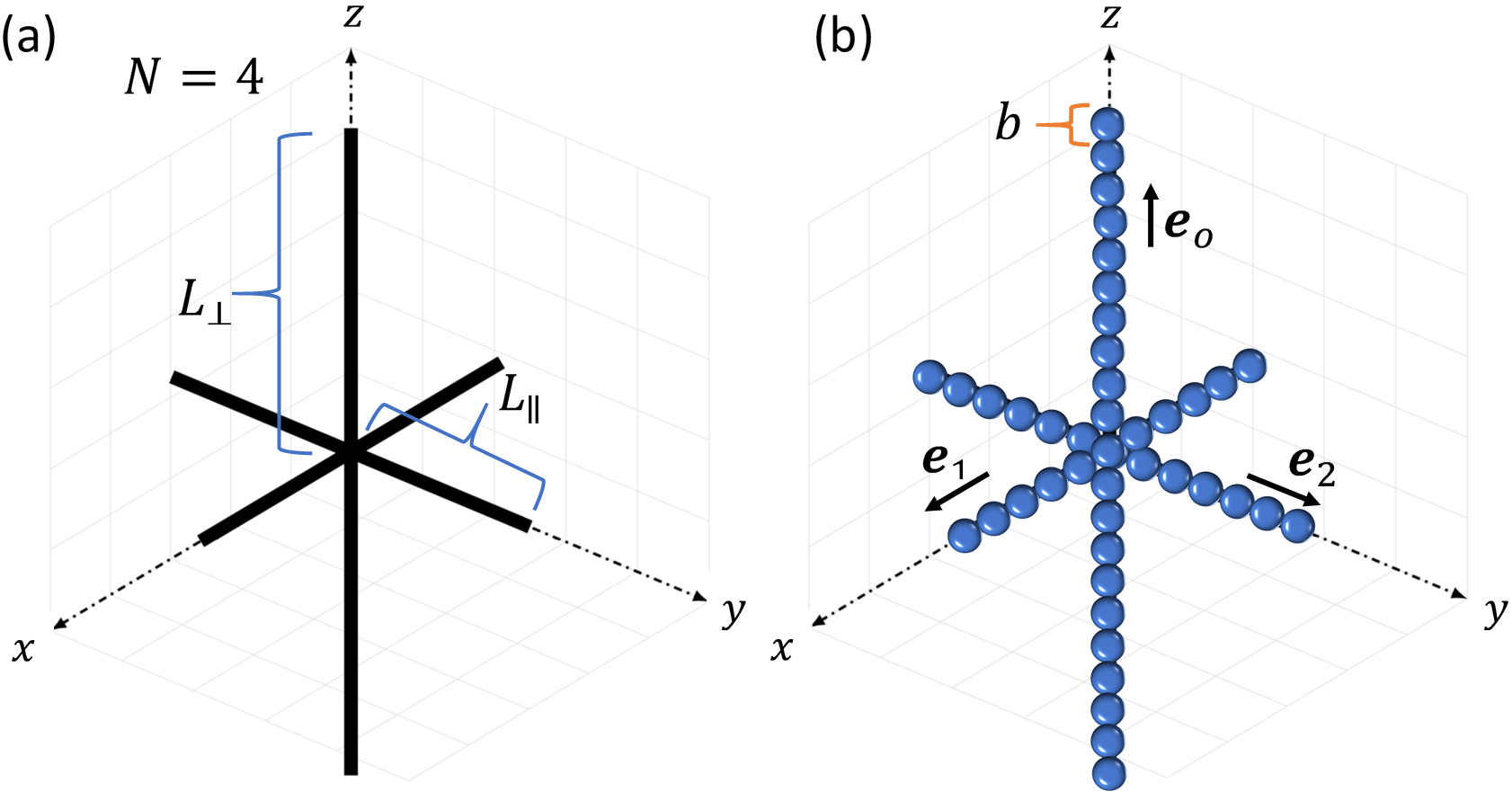}}
\caption{\label{fig:schematic} (a) One of the wire-frame particles considered, with $N=4$ in-plane legs and unequal lengths of the in and out of plane legs, $L_{\parallel} \neq L_{\bot}$ (b) The `shish-kebab' procedure as applied to the shape from panel (a). The wire-frame is replaced by spheres with diameter $b$ placed along each leg of the shape. The unit vectors $\textbf{e}_{l}$ associated with some of the legs are also indicated.}
\end{figure}

\paragraph{Wire Frame Shapes:}\label{sec:wireframeshapes}
We consider shapes comprised of rigid, thin rods or legs. Each leg is indexed by $l$ and has a different length $L_l$. We call these shapes `wire frames'. We only consider shapes where legs meet at one point, see Fig.(\ref{fig:schematic}), but the formulae can be easily modified when this is not the case. 

These particles provide a simple way to test our predictions and construct shapes of different symmetries within the same framework. They are also presented as a model for recently synthesised DNA nano-particles \cite{Xing2018MicrorheologyHydrogels}. Since double stranded DNA is very rigid with persistance length of $\sim 390$\r{A} \cite{Gross2011QuantifyingTension} and a typical aspect ratio $\sim 20$, the approximation of rigidity and large aspect ratios should be appropriate. We construct a formulation, based on Onsager's variational principle\cite{Onsager1931ReciprocalI.,Onsager1931ReciprocalII.,Doi2011OnsagersMatter,Doi2013SoftPhysics}, to describe these shapes in general, into which any given shape may be specified.

The principle states that the linear and angular velocities of the particle are those which extremise the Rayleighian, $\mathcal{L}$, of the system. The Rayleighian is defined as, $\mathcal{L} = \dot{F} + \frac{1}{2}\Phi$, where $\dot{F}$ is the time derivative of the Helmholz free energy and $\Phi$ is the energy dissipation function. 

To determine the Rayleighian we use a standard `Shish-Kebab' approach \cite{Doi1986TheDynamics}. We place $N_l$ spherical beads of radius $b$ along each leg of the shape, such that they just touch, i.e. the length of the $l^{th}$ leg is $ L_l = N_l b$. Fig.(\ref{fig:schematic}b) shows this for a particular wire frame shape. The position vector and velocity of the $n^{th}$ bead on leg $l$ are written as
\begin{equation}
\label{positions}
    \textbf{r}_n^l = \textbf{r} + n b \textbf{e}_{l} \ \ , \  \ \textbf{v}_n^l = \textbf{v} + n b \ \bm{\Omega} \times \textbf{e}_{l},
\end{equation}
where $n \in [0,N_{l}]$ and the unit vector $\textbf{e}_l$ points along the leg.

The velocities of all the beads are in general coupled together by hydrodynamic interactions. This coupling is difficult to handle exactly, but a simple approximation can be made. We assume that the rods are very long compared to their width, such that the majority of beads on different legs are very far apart. This allows the hydrodynamic interaction between beads on different legs to be neglected, which becomes an accurate approximation in the limit of infinite aspect ratio, $L_l/b$.

Using a textbook procedure \cite{Doi2013SoftPhysics,Doi1986TheDynamics} the Rayleighian for a general wire frame shape can be determined, providing a complete description of its behaviour \cite{SeeDerivation.}. We focus on the linear viscoelastic properties, so we only consider terms containing the angular velocity. The relevant terms in the Rayleighian are
\begin{equation}
\label{rayleighianangvel}
\begin{split}
&\mathcal{L} = \big\langle \bm{\Omega} \cdot \bm{\mathcal{R}}  k_{B} T \log \psi \big\rangle + \frac{1}{2} \Big\langle \sum_{l} \lambda_{l} (\bm{\Omega} \times \textbf{e}_l) \cdot (\textbf{v} - \kappa \cdot \textbf{r}) \Big \rangle \\
& + \frac{1}{2}\Big\langle \sum_{l} \mu_{l} \ (\bm{\Omega} \times \textbf{e}_l)^2 - 2 \mu_{l} \ \bm{\Omega} \cdot (\textbf{e}_l \times \kappa \cdot \textbf{e}_l) \Big\rangle + (\cdots),
\end{split}
\end{equation}
where we have defined the friction constants $\lambda_{l}$ and $\mu_{l}$ as
\begin{equation}
\label{rotfric}
 \lambda_l = \frac{4\pi \eta_s L_l^2}{\log (L_l / b)} \ \ \text{and} \ \ \mu_{l} = \frac{8\pi \eta_s L_l^3}{3 \log (L_l / b)}.
\end{equation}
By appropriately choosing the unit vectors, $\textbf{e}_l$, we can describe any wire frame shape.

We specifically consider shapes comprised of $N$ evenly spaced, co-planar legs of equal length with two anti-parallel legs pointing orthogonally out of plane, as shown in Fig.(\ref{fig:schematic}). The lengths of the in and out of plane legs are $L_{\parallel}$ and $L_{\bot}$ respectively. The in plane legs are separated by an angle $\phi = 2\pi/N$. 
The unit vectors parallel to the legs in plane are denoted, $\textbf{e}_l = \cos(l\phi) \textbf{u}_1 - \sin(l\phi) \textbf{u}_l$ for $ 1 \leq l \leq N $. For the out of plane legs, $\textbf{e}_{-1,0} = \pm \textbf{u}_3$ \footnote{We could have defined only one vector for these legs, but this notation is more consistent across a wider range of shapes.}. 

\begin{figure}
{\includegraphics[width=12cm]{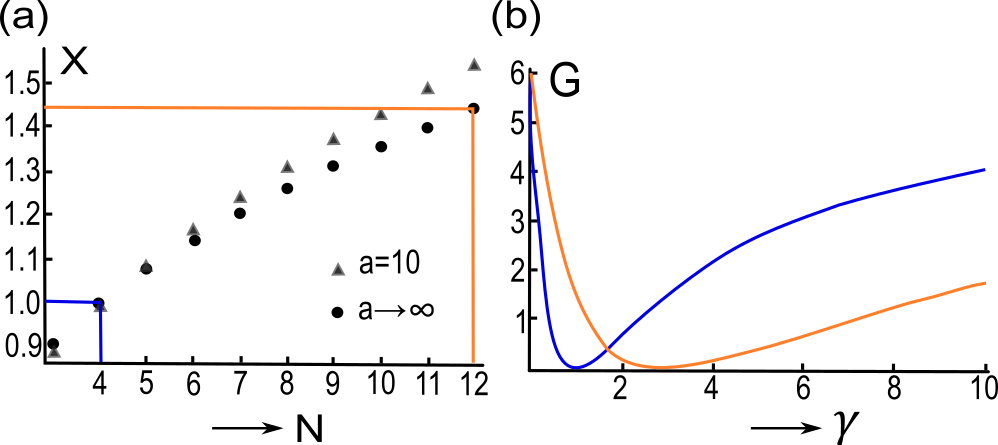}}
\caption{\label{fig:aspectratio} (a) The required ratio of leg lengths, $x = L_{\bot}/L_{\parallel}$, for the elasticity to vanish plotted as a function of the number of in-plane legs, $3 \leq N \leq 12$, for aspect ratios, $a= L_{\bot}/b=10$ (triangles) and $a \to \infty$ (circles). The (dark) blue and (light) orange lines indicate the solutions for $N=4$ and $N=12$ respectively. Panel (b) shows the elastic modulus, $G$ a function of $\gamma \approx L_{\bot}^3/L_{\parallel}^3$. The blue curve corresponds to $N=4$ and the orange $N=12$.}
\end{figure}

According to Onsager's principle \cite{Onsager1931ReciprocalI.,Onsager1931ReciprocalII.,Doi2013SoftPhysics,Doi2011OnsagersMatter}, the components of the stress tensor are given by, $\sigma_{i j} = \partial \mathcal{L}^{*}/ \partial \kappa_{i j}$, where $\mathcal{L}^{*}$ is the Rayleighian (\ref{rayleighianangvel}) evaluated at the extremum value of $\bm{\Omega}$. We find that the Brownian stress tensor as a function of time is given by \cite{SeeDerivation.},
\begin{equation}
(S_{B})_{ij} = n k_B T \ G \int_{-\infty}^{t} dt' \ e^{-(t - t')/\tau } \kappa_{ij}(t'),
\end{equation}
where $\tau$ is a decay timescale and $G$ is the elastic modulus
\begin{equation}
 G = 6 \bigg(\frac{N- 4 \gamma}{N + 4 \gamma} \bigg)^2, 
\end{equation}
with $\gamma \equiv \mu_{\bot}/\mu_{\parallel}$. The friction constants, $\mu_{\parallel}$ and $\mu_{\bot}$, are calculated from (\ref{rotfric}) for the in and out-of-plane legs respectively. This modulus is plotted in Fig.(\ref{fig:aspectratio}b) for $N=4$ and $N=12$.

The elastic modulus has a minimum of zero when $\gamma = N/4 $. This means that when the ratio of leg lengths, $x =L_{\bot}/ L_{\parallel}$, satisfies the transcendental equation,
\begin{equation}
\label{exactratio}
    4 x^3 \log( a / x ) =  N \log a,
\end{equation}
the suspension is purely viscous. The aspect ratio $a$ is defined as $L_{\bot}/b$. This recovers the expected result; a dilute suspension of symmetric cross shapes with $N=4$ and $L_{\bot}=L_{\parallel}$ (equivalent to an octahedron or a cube) has a purely viscous linear stress response. 

A plot of the solutions to this equation for $3\leq N \leq 12$ and $a=10$ is shown in Fig.(\ref{fig:aspectratio}a, triangles), we also show the solutions in the limit $a \to \infty$ (circles), which allows $\gamma$ to be approximated as $(L_{\bot}/L_{\parallel})^3$.

It is intriguing that we can engineer the elasticity to vanish for \textit{any} $N$ by choosing the right ratio, $x$. For instance, when $N=3$ the particle has the symmetry of a trigonal bi-pyramid of variable height. The symmetry group for such an object does not satisfy the conditions given previously, yet when the ratio of lengths is chosen appropriately the elasticity still vanishes, in the small $\kappa$ regime.

\begin{figure}
{\includegraphics[width=12cm]{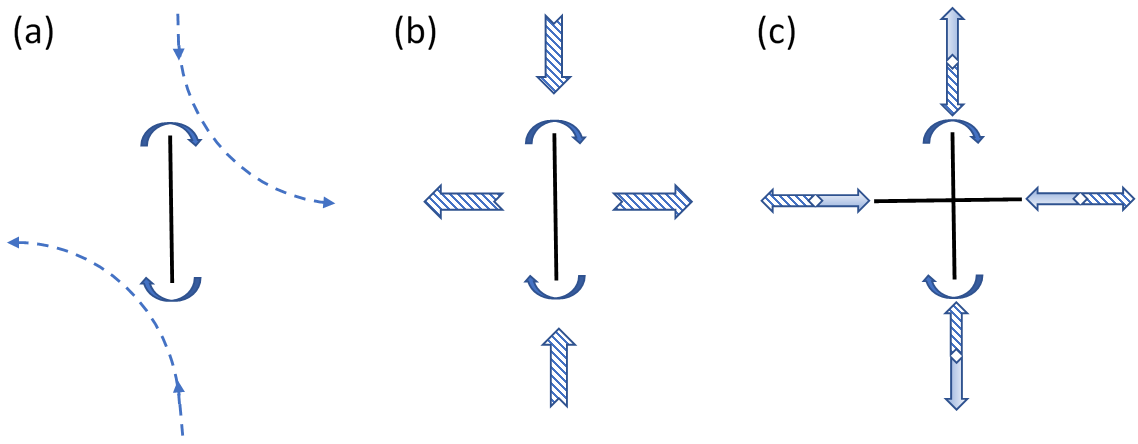}}
\caption{\label{fig:stresslets} (a) The rotation of the black rod about the axis into the page, shown by the solid arrows, causes a fluid packet to move along the streamline represented by the dotted blue lines. (b) This results in a net flow towards the ends of the rod and away from the broad-sides, characteristic of the stresslet singularity. (c) When the cross rotates, we sum the flows from the constituent rods. The flow induced by the horizontal rod is shown by the shaded arrows, and the striped arrows for the vertical rod. If the rods are the same length then the magnitudes of these two flows are equal and they cancel.}
\end{figure}

This phenomenon, while not explained by a simple symmetry argument, can be physically understood by considering the stresslet produced by the rotation of the particle. When a rod rotates about an axis perpendicular to its length in either direction, the surrounding fluid flows towards either end of the rod but away from the broad-side of the rod, as shown in Figs.(\ref{fig:stresslets}a) \& (\ref{fig:stresslets}b). The resulting flow is typical of the stresslet singularity, whose magnitude depends on the length of the rod. 

Fig.(\ref{fig:stresslets}c) shows the fluid flow induced by the rotation of a planar cross, formed of two rods which perpendicularly bisect each other. If the constituent rods are the same length, then the stresslet flows cancel each other in the plane of the shape. This construction may be applied to the $N=4$, $\gamma=1$ wire frame. Shapes where $N \neq 4$ can be engineered to produce no stresslet when they rotate by appropriately tuning the magnitude of the stresslet produced by the out of plane rod to cancel that produced by the in plane shape.
\paragraph{Conclusions:} We have discussed the origins of non-Newtonian rheology for dilute suspensions of rigid particles and determined sufficient conditions the shape must satisfy for a dilute suspension to be Newtonian. To have a purely Newtonian response for all strain sizes the symmetry group of the shape must have irreducible representations of degree 3 and degree 5, whereas in the regime of small strains the symmetry group only needs an irreducible representation of degree 3 for Newtonian behaviour. This allows for simple classification of suspensions without the need for detailed calculation.

We also developed a framework for studying the rheology of wire-frame particles constructed from thin, rigidly connected rods using Onsager's variational principle. This was used to demonstrate the vanishing elasticity for octahedral and cubic shapes in the linear regime, as well as find a set of bi-pyramidal shapes which, despite their symmetry group not satisfying the appropriate condition, have Newtonian dilute suspensions for small strains. This is physically explained in terms of the stresslets produced by the rotation of each constituent rod. 

The study of wire frame shapes has relevance for the design of DNA nanostar suspensions where understanding which particle shapes lead to a particular rheological response is very important.

\paragraph{Acknowledgements:} We are grateful to Prof. Daan Frenkel for insightful conversations and valuable comments on the manuscript. DAK acknowledges financial support from the UK Engineering and Physical Sciences Research Council.

\bibliography{ParticleShapesLeadingtoNewtonianDiluteSuspensions}

\end{document}


\preprint{APS/123-QED}

\title{\underline{Supplementary Material} to ``Particle shapes leading to Newtonian dilute suspensions"}


\author{David A. King} \email{dak43@cam.ac.uk}
\affiliation{%
Cavendish Laboratory, University of Cambridge, J. J. Thomson Ave., Cambridge CB3 0HE, UK
}%
\author{Masao Doi}%
\affiliation{%
Centre of Soft Matter and its Applications, Beihang University, Beijing 100191, China 
}%
\author{Erika Eiser} \email{ee247@cam.ac.uk}
\affiliation{%
Cavendish Laboratory, University of Cambridge, J. J. Thomson Ave., Cambridge CB3 0HE, UK
}%

\date{\today}
             
\maketitle

In this document, we present the details of the derivation of the stress in a dilute suspension of wire frame frame particles which was presented in the main text. The derivation is based on Onsager's variational principle \cite{Onsager1931ReciprocalI.,Onsager1931ReciprocalII.,Doi2013SoftPhysics} and begins by determining the Rayleighian, from which we determine the angular velocity of the particle. This is enough to determine the stress in the system using the methods in Makino and Doi \cite{Makino2004ViscoelasticityShape}. \\

The Rayleighian of the system is defined as in the main text,
\begin{equation}
\label{rayleighiangen}
\mathcal{L} = \dot{F} + \frac{1}{2}\Phi,
\end{equation}
where $\dot{F}$ is the time derivative of the free energy and $\Phi$ is the energy dissipation function. 

The Free Energy is written in standard from in terms of the distribution function $\psi( \textbf{u}_1, \textbf{u}_2; \textbf{r})$,
\begin{equation}
\label{freeenergygeneral}
 F = \int d\textbf{r} \ d\textbf{u}_1 d\textbf{u}_2 \ k_{B} T \psi \log \psi + \psi U,
\end{equation}
where $U$ is the external potential. Taking the time derivative of this expression we find that, 
\begin{equation}
\label{Fdot}
\dot{F} = \int d\textbf{r} \ d\textbf{u}_1 d\textbf{u}_2 \ k_{B} T \ \dot{\psi} \ (\log \psi +1 + \beta U),
\end{equation}
with $\beta = 1/k_{B}T$.

The orientational distribution function $\psi$ must satisfy the standard contintuity equation, 
\begin{equation}
\label{ctyeqn}
\frac{\partial \psi}{\partial t} = - \frac{\partial}{\partial \textbf{r}} \cdot (\textbf{v} \psi) - \bm{\mathcal{R}} \cdot (\bm{\Omega} \psi),
\end{equation}
where $\textbf{v}$ and $\bm{\Omega}$ are the linear and angular velocities of the particle as in Eqn.(1) of the main text. This is substituted into (\ref{Fdot}) and integrated by parts to obtain,
\begin{equation}
\dot{F} = \int d\textbf{r} \ d\textbf{u}_1 d\textbf{u}_2 \ \psi \ \left( \bm{\Omega} \cdot \bm{\mathcal{R}} + \textbf{v} \cdot \frac{\partial}{\partial \textbf{r}} \right) \ (k_{B} T \ \log \psi+ U) = \bigg\langle\left( \bm{\Omega} \cdot \bm{\mathcal{R}} + \textbf{v} \cdot \frac{\partial}{\partial \textbf{r}} \right) \ (k_{B} T \ \log \psi+ U) \bigg\rangle.
\end{equation}
To determine the energy dissipation function, we consider a standard `Shish-Kebab' approach as outlined in the main text. Here we place $N_l$ spherical beads of radius $b$ along each leg of the shape, indexed by $l$. The beads should be placed such that they just touch, meaning that the length of the $l$th leg is $ L_l = N_l b$. The position vector of the $n^{th}$ on leg $l$ is then written,
\begin{equation}
\label{positions}
    \textbf{r}_n^l = \textbf{r} + n b \textbf{e}_{l},
\end{equation}
and the velocity of this bead therefore is 
\begin{equation}
\label{velocities}
\textbf{v}_n^l = \textbf{v} + n b \ \bm{\Omega} \times \textbf{e}_{l}
\end{equation}
When each bead moves through the surrounding fluid with this velocity it experiences a viscous drag force, $\textbf{F}_n^{l}$. This leads to the power dissipated per bead, $\textbf{F}_n^{l} \cdot (\textbf{v}_n^l - \kappa \cdot \textbf{r}_n^{l})$, which is then summed over all of the beads and averaged over orientations to obtain the energy dissipation function. 

The viscous drag force can be calculated in terms of the velocity of the other beads and the inverse mobility matrix, 
\begin{equation}
\textbf{F}_n^{l} = \sum_{l'=0}^{N}\sum_{m=0}^{N_{l'}} (H^{-1})_{nm}^{l,l'} \cdot (\textbf{v}_m^{l'} - \kappa \cdot \textbf{r}_m^{l'}),   
\end{equation}
where $(H^{-1})_{nm}^{l,l'}$ is the inverse mobility matrix and expresses the hydrodynamic coupling of the $m^{th}$ bead on leg $l'$ and the $n^{th}$ bead on leg $l$. As discussed in the main text, we simplify this by neglecting the coupling between beads on different legs of the shape. This amounts to assuming that the rods are very long compared to their width such that the majority of beads on different legs are very far apart, which becomes accurate in the limit of infinite aspect ratio $L_l/b$. In this approximation the inverse mobility matrix becomes,
\begin{equation}
    (H^{-1})_{n,m}^{l,l'} \approx \delta_{l,l'} (H^{-1})_{n,m}^l.
\end{equation}
Since the beads are all spherical, we can determine $(H^{-1})_{n,m}^l$ from the Oseen Tensor in a standard way \cite{Kim2005MicrohydrodynamicsApplications,Happel1973LowMedia.,Doi1986TheDynamics},
\begin{equation}
    (H^{-1})_{n,m}^{l,l'} \approx  \delta_{l,l'} (h^{-1})_{n,m} (\mathbb{I} - \frac{1}{2} \textbf{e}_l \textbf{e}_l),
\end{equation}
where the constants $(h^{-1})_{n,m}$ are determined by the relation,
\begin{equation}
    \sum_{p=0}^{N_p}  \frac{1}{8 \pi \eta_s |p-m|} (h^{-1})_{n,p}= \delta_{n,m}.
\end{equation}
Using these expressions, the energy dissipation function becomes
\begin{equation}
\label{EnergyDiss}
\Phi = \Big\langle\sum_{n,m;l}  (h^{-1})_{n,m} (\textbf{v}_n^l - \kappa \cdot \textbf{r}_n^l) \cdot  (\mathbb{I} - \frac{1}{2} \textbf{e}_l \textbf{e}_l) \cdot (\textbf{v}_m^{l} - \kappa \cdot \textbf{r}_m^{l})  \Big\rangle.  
\end{equation}

Substituting the expressions for the positions of the beads (\ref{positions}) and the velocities (\ref{velocities}) into (\ref{EnergyDiss}) we find Rayleighian for a general wireframe shape, 
\begin{equation}
\label{RayleighianGeneral}
\begin{split}
\mathcal{L} &= \bigg\langle\left( \bm{\Omega} \cdot \bm{\mathcal{R}} + \textbf{v} \cdot \frac{\partial}{\partial \textbf{r}} \right) \ (k_{B} T \ \log \psi+ U) \bigg\rangle \\
&+ \bigg\langle\sum_{l}\sum_{n,m = 0}^{N_{l}} (h^{-1})_{nm} (\textbf{v} - \kappa \cdot \textbf{r}) \cdot (\mathbb{I} - \frac{1}{2} \textbf{e}_l \textbf{e}_l) \cdot (\textbf{v} - \kappa \cdot \textbf{r})\bigg\rangle \\
&+ \bigg\langle\sum_{l}\sum_{n,m = 0}^{N_{l}} n m b^2 (h^{-1})_{nm} [(\bm{\Omega} \times \textbf{e}_l)- \kappa \cdot \textbf{e}_l] \cdot  (\mathbb{I} - \frac{1}{2} \textbf{e}_l \textbf{e}_l) \cdot [(\bm{\Omega} \times \textbf{e}_l)- \kappa \cdot \textbf{e}_l]\bigg\rangle \\
&+ 2 \bigg\langle\sum_{l}\sum_{n,m = 0}^{N_{l}} n (h^{-1})_{nm} (\textbf{v} - \kappa \cdot \textbf{r}) \cdot  (\mathbb{I} - \frac{1}{2} \textbf{e}_l \textbf{e}_l) \cdot [(\bm{\Omega} \times \textbf{e}_l)- \kappa \cdot \textbf{e}_l]\bigg\rangle.
\end{split}
\end{equation}
To simplify notation, we identify the three constants,
\begin{equation}
\nu_l = \sum_{n,m=0}^{N_l} (h^{-1})_{nm} \ \ , \ \ \mu_l = \sum_{n,m=0}^{N_l} nm b^2 (h^{-1})_{nm} \ \ , \ \ \lambda_l = \sum_{n,m=0}^{N_l} nb(h^{-1})_{nm}.
\end{equation}
Using the techniques in Appendix 8.I of \cite{Doi1986TheDynamics}, these can be evaluated to give the expressions in Eqn.(12) of the main text,
\begin{equation}
\nu_l = \frac{8 \pi \eta_s L_l}{\log (L_l / b)} \ \ , \ \ \mu_l = \frac{8 \pi \eta_s L_l^3}{3\log (L_l / b)}  \ \ , \ \ \lambda_l = \frac{4 \pi \eta_s L_l^2}{\log (L_l / b)}. 
\end{equation}
This, in principle, provides a complete description of the behaviour of a dilute suspension of general wire frames. From now on, we restrict our attention to the specific class of shapes discussed in the main text, where the vectors $\textbf{e}_l$ are defined as,
\begin{equation}
\label{vecs}
\begin{split}
&\textbf{e}_l = \cos(l\phi) \textbf{u}_1 - \sin(l\phi) \textbf{u}_l \ \ , \ \ 1 \leq l \leq N \\
&\textbf{e}_{-1,0} = \pm \textbf{u}_3,
\end{split}
\end{equation}
with $\phi = 2\pi/N$.

Using these definitions it is straightforward to show that the third term in the Rayleighian (\ref{RayleighianGeneral}) is zero because it includes the sums, 
\begin{equation}
\begin{split}
&\sum_{n = 0}^{N-1} \cos^3 \left(\frac{2 \pi n}{N}\right) = \sum_{n = 0}^{N-1} \sin^3 \left(\frac{2 \pi n}{N}\right) = 0 \ \ ; \ \ N \geq 3 \\
&\sum_{n = 0}^{N-1} \sin \left(\frac{2 \pi n}{N}\right) \cos^2 \left(\frac{2 \pi n}{N}\right) = \sum_{n = 0}^{N-1} \cos \left(\frac{2 \pi n}{N}\right) \sin^2 \left(\frac{2 \pi n}{N}\right) = 0 \ \ ; \ \ N \geq 3.
\end{split}
\end{equation}

According to Onsager's Principle, the linear and angular velocities appearing in the continuity equation (\ref{ctyeqn}) are those which maximise the Rayleighian. It is easy to determine that the linear velocity satisfies,
\begin{equation}
\label{linvel}    
\sum_{l} \nu_l (\mathbb{I} - \frac{1}{2} \textbf{e}_l \textbf{e}_l) \cdot (\textbf{v} - \kappa \cdot \textbf{r}) = - \frac{\partial}{\partial \textbf{r}} (k_{B} T \log \psi + U),
\end{equation}
and the angular velocity satisfies,
\begin{equation}
\label{angvel}
\sum_{l}\ \mu_l \ \textbf{e}_l \times (\bm{\Omega} \times \textbf{e}_l) = - \bm{\mathcal{R}} (k_{B} T\log \psi +U) + \sum_{l} \mu_l \ \textbf{e}_l \times \kappa \cdot \textbf{e}_l.
\end{equation}
In our case we focus on the viscoelastic properties so we only consider the angular velocity and set the external potential to zero. By using the definitions (\ref{vecs}) as well as the definitions of $\mu_{\parallel}$ and $\mu_{\bot}$ as given in the text, we can write the equation for the angular velocity explicitly; 
\begin{equation}
\begin{split}
&\left[\sum_{l=0}^{N-1} \mu_{\parallel}
\begin{pmatrix}
    1-c^2 & - s c & 0 \\
    - s c & 1-s^2 & 0 \\
    0 & 0 & 1
\end{pmatrix}
+ 2\mu_{\bot}\begin{pmatrix}
    1 & 0 & 0 \\
    0 & 1 & 0 \\
    0 & 0 & 0
\end{pmatrix} \right] \bm{\Omega} = -\bm{\mathcal{R}} k_{B}T \log \psi + \mu_{\parallel} \sum_{l=0}^{N-1}\begin{pmatrix}
    s^2 \kappa_{32} \\
    -c^2 \kappa_{31} \\
    c^2 \kappa_{21} - s^2 \kappa_{12}
\end{pmatrix}
+ 2\mu_{\bot} \begin{pmatrix}
    -\kappa_{23} \\
    \kappa_{13} \\
    0
\end{pmatrix},
\end{split}    
\end{equation}
where we have defined $c=\cos l\phi$ and $s= \sin l\phi$. Using the sums,
\begin{equation}
\begin{split}
&\sum_{n = 0}^{N-1} \cos^2 \left(\frac{2 \pi n}{N}\right) = \sum_{n = 0}^{N-1} \sin^2 \left(\frac{2 \pi n}{N}\right) = \frac{N}{2} \ \ ; \ \ N \geq 3 \\
&\sum_{n = 0}^{N-1} \sin \left(\frac{2 \pi n}{N}\right) \cos \left(\frac{2 \pi n}{N}\right) = 0 \ \ ; \ \ N \geq 3 .
\end{split}   
\end{equation}
By solving this equation and comparing to Eqn.(1) of the main text, we find the following forms for the mobility tensors $c$ and $\tilde{h}$.
\begin{equation}
\label{coeffsstar}
\begin{split}
& c_1 = c_2  = \frac{1}{\mu_{\parallel}(N/2 + 2 \gamma)} \ \ c_3 = \frac{1}{\mu_{\parallel} N } \\
& \Tilde{h}_{231} = -\Tilde{h}_{132} = \frac{N/2}{N/2 + 2 \gamma} \ \ , \ \ \Tilde{h}_{123} = -\Tilde{h}_{213} = \frac{2 \gamma}{N/2+2 \gamma} \\
& \Tilde{h}_{312} = -\Tilde{h}_{321} = \frac{1}{2}.
\end{split}
\end{equation}
All other components of $\tilde{h}$ are zero and $\gamma \equiv \mu_{\bot} / \mu_{\parallel}$. 

We may now determine the elastic stress by following the steps laid out in Makino and Doi \cite{Makino2004ViscoelasticityShape}. There they show that the elastic stress, $\sigma_E$, as a function of time is given by, 
\begin{equation}
    \sigma_E(t) = n k_{B} T \sum_{i=1}^{5} \int_{-\infty}^{t} dt' \ G_i e^{-(t-t')/\tau_i} \kappa(t'), 
\end{equation}
where the constants, $G_i$, are elastic moduli corresponding to the different decay time scales, $\tau_i$. 

For our purposes, it is only necesary to determine the moduli, $G_i$, which are in general given by,
\begin{equation}
\begin{split}
&G_1 = \frac{4\left[(c_2-c_3)(\tilde{h}_{231}-\tilde{h}_{321})+(c_1-c_2+D)(\tilde{h}_{312}-\tilde{h}_{132})+(c_3-c_1-D)(\tilde{h}_{123}-\tilde{h}_{213})\right]^2}{(c_2-c_3)^2+(c_1-c_2+D)^2+(c_3-c_1-D)^2}\\
&G_2 =  \frac{4\left[(c_2-c_3)(\tilde{h}_{231}-\tilde{h}_{321})+(c_1-c_2-D)(\tilde{h}_{312}-\tilde{h}_{132})+(c_3-c_1+D)(\tilde{h}_{123}-\tilde{h}_{213})\right]^2}{(c_2-c_3)^2+(c_1-c_2-D)^2+(c_3-c_1+D)^2}\\
&G_3 = 2\left(\tilde{h}_{232}-\tilde{h}_{322}+\tilde{h}_{311}-\tilde{h}_{113}\right)^2\\
&G_4 = 2\left(\tilde{h}_{313}-\tilde{h}_{133}+\tilde{h}_{122}-\tilde{h}_{221}\right)^2\\
&G_5 =  2\left(\tilde{h}_{121}-\tilde{h}_{211}+\tilde{h}_{233}-\tilde{h}_{332}\right)^2.
\end{split}    
\end{equation}
Where $D^2 = c_1^2 + c_2^2 + c_3^2 - c_1 c_2 - c_1 c_3 -c_2 c_3$.

From the forms of $c$ and $\tilde{h}$ in (\ref{coeffsstar}) it is evident that the final three moduli vanish, $G_3=G_4=G_5=0$. After some tedious algebra we find also find that $G_2=0$. Thus, we are left with $G_1$, which after more manipulation becomes the expression given in Eqn(14) of the main text,
\begin{equation}
 G = 6 \bigg(\frac{N- 4 \gamma}{N + 4 \gamma} \bigg)^2.     
\end{equation}

\bibliography{SuppMatt}